\documentclass{article}
\usepackage{spconf,amsmath,graphicx}
\usepackage{bm}
\usepackage{amsfonts}

\title{CROSS-MODAL SPECTRUM TRANSFORMATION NETWORK FOR ACOUSTIC SCENE CLASSIFICATION}
\name{Yang~Liu$^1$,  Alexandros Neophytou$^2$, Sunando Sengupta$^2$, Eric Sommerlade$^2$ \thanks{This work is done by Yang Liu as intern researcher in Microsoft. The code and dataset are published in https://git.io/JtK3I.}}
\address{$^1$ Department of Electrical and Electronic Engineering, University of Surrey, UK\\
$^2$ Microsoft Corporation, Reading, UK\\ 
E-mail: yangliu@surrey.ac.uk; Alexandros.Neophytou, Sunando.Sengupta, Eric.Sommerlade@microsoft.com}

\begin{document}
%\ninept
%
\maketitle
\begin{abstract}

Convolutional neural networks (CNNs) with log-mel spectrum features have shown promising results for acoustic scene classification tasks. However, the performance of these CNN based classifiers is still lacking as they do not generalise well for unknown environments. To address this issue, we introduce an acoustic spectrum transformation network where traditional log-mel spectrums are transformed into imagined visual features (IVF). The imagined visual features are learned by exploiting the relationship between audio and visual features present in video recordings. An auto-encoder is used to encode images as visual features and a transformation network learns how to generate imagined visual features from log-mel. Our model is trained on a large dataset of Youtube videos. We test our proposed method on the scene classification task of DCASE and ESC-50, where our method outperforms other spectrum features, especially for unseen environments. 

\end{abstract}

\section{Introduction}

Acoustic scene classification is a popular task that aims to classify scenes with audio recordings into different classes. It has many applications in surveillance of  sound event detection \cite{kong2018dcase}. However, it is difficult to classify acoustic scenes for unseen environments and unknown recording equipment due to the lack of appropriate labelled data-sets. Current state of the art approaches uses various augmentation methods to address this issue. However, over-fitting happens frequently when a large number of augmentation data is used \cite{lemley2017smart}.

In this work, we introduce an acoustic scene transformation algorithm to address the unseen environment problem for acoustic scene classification. The goal is to generate imagined visual features from short audio recordings and to minimise the divergence between the imagined visual features and real visual features. Real visual features are calculated from images obtained from training videos using an auto-encoder. A transformation network is used to learn the mapping between log-mel \cite{davis1980comparison} features and visual features. In order to decrease the complexity of learning, a vector quantisation based codebook is applied and shared between the auto-encoder and the transformation network. Since the videos used for training are recorded using different equipment in various cities, the imagined visual features are more robust than the traditional features for unseen cities. For classifying the acoustic scene, a classifier is used to predict the category of audio recordings based on the imagined visual features. The auto-encoder and transformation network are trained on the Youtube 8M dataset \cite{abu2016youtube}. The classifier is trained on the target dataset, such as DCASE \cite{mesaros2019acoustic} and ESC-50 \cite{piczak2015esc}. 

\begin{figure*}[tb]
    \centering
    \includegraphics[width=0.95\linewidth]{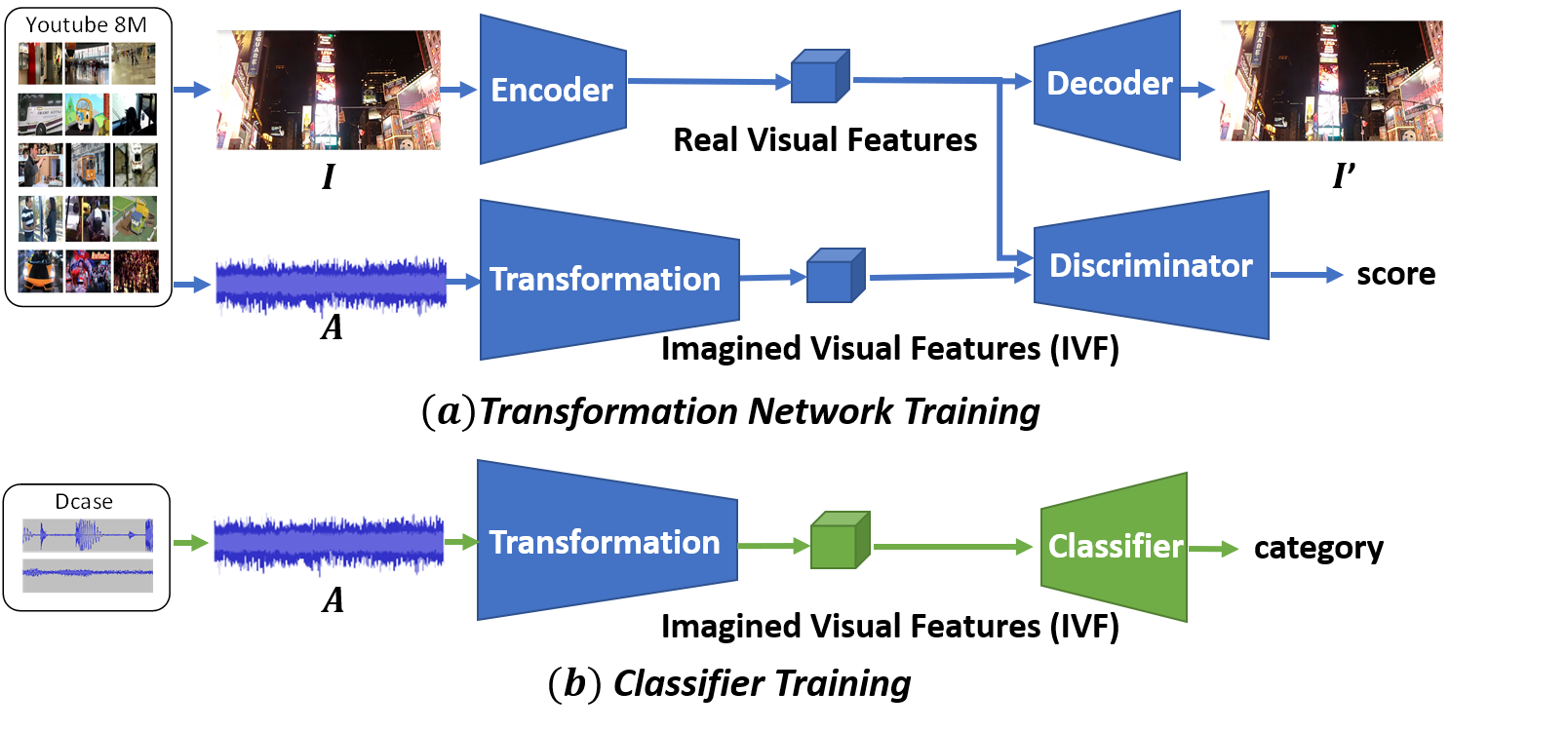} 
    \caption{Proposed model and training pipeline. In sub-figure (a), Audio-visual pairs of the Youtube 8M dataset are used to train the encoder, decoder and transformation network (Blue lines). In sub-figure (b), The audio recording of DCASE dataset is used to train a classifier (Green lines) which predicts the category of the audio recording when the transformation network is fixed. }
    \label{fig:pipeline}
\end{figure*}

\section{Related Work}

\subsection{Acoustic scene classification}

Many audio classification methods use deep learning networks, such as convolutional neural networks (CNNs) \cite{koutini2019cp} and recurrent neural networks (RNNs) \cite{kong2018dcase}. Suh et al. \cite{suh2020designing} introduce trident ResNet with the categorical focal loss to address unseen devices problem. Hu et al. \cite{hu2020device} propose a novel two-stage network with two CNNs, classifying the acoustic input according to high-level categories, and then low-level categories, respectively. Koutini et al. \cite{koutini2019cp} propose a novel frequency-aware CNN layer and adaptive weighting for addressing the noisy labels problem, which achieves the top accuracy for unseen cities in the 2019 DCASE challenge. These CNN based methods commonly use the log-mel spectrum of audio recordings as input and output the class probability of an acoustic scene in either frame-level or video-level. However, the spectral content of recordings belonging to the same class can vary greatly between different cities, especially when the recording equipment is not matched \cite{koutini2019cp}.  

\subsection{Audio-visual cross-modal learning}

Audio-visual data offer a wealth of resources for knowledge transfer between different modalities \cite{aytar2016soundnet}. Our work is closely related to the reconstruction from audio recordings. For example, Oh et al. \cite{oh2019speech2face} and Athanasiadis et al. \cite{athanasiadis2019audio} have used speech to reconstruct the speakers' face. Chen et al. \cite{chen2017deep} Hao et al. \cite{hao2018cmcgan} use conditional GANs and cycle GAN to achieve the cross-modal audio-visual generation of musical performances. However, these works focus on special cases, such as faces or musical instruments, where the visual object is located at the centre of the images. For less constrained test data, the reconstructed images may be noisy, due to large variation.  Arandjelovic et al. \cite{arandjelovic2017look} propose a generic audio-visual model to classify if a given video frame and a short audio clip correspond to each other, where the fusion layers only select the common features of the recording and images. Aytar et al. \cite{aytar2016soundnet} propose a student-teacher training procedure where a pre-trained visual recognition model transfers the knowledge from the visual modality to the sound modality. Compared to Aytar et al. \cite{aytar2016soundnet}, our proposed method our proposed method does not need any pre-trained visual recognition model as the teachers of the audio model. 

\section{Spectrum transformation Network for Classification }

In this section, we introduce the spectrum transformation network and discuss how to train its objective function along with implementation details. The transformation network with classifier is shown in Fig. \ref{fig:pipeline}. 

\subsection{Transformation network}

The acoustic spectrum transformation network aims to learn a mapping function between acoustic features and visual features from unlabelled videos. Although unlabelled videos do not contain the category information, they provide a large number of audio-visual pairs $\{\bm{I}, \bm{A}\}$, where $\bm{A}$ is the log-mel of a fixed short length audio recording and $\bm{I}$ is randomly selected frame of the corresponding video. The main challenge for training the acoustic spectrum transformation network is that the distributions of the audio samples and visual samples are different. 
%and their Jensen–Shannon (JS) divergence \cite{manning1999foundations} is large

The visual features of $\bm{I}$ are calculated by an auto-encoder based on VQ-VAE \cite{van2017neural}, which is a reconstruction system. As shown in Fig. \ref{fig:pipeline}, the audo-encoder has an encoder and an decoder where the encoder maps the input image $\bm{I}$ to a sequence of discrete latent variables and the decoder reconstruct a image ${\bm{I}}'$ from the latent variables. A codebook is applied between the encoder and decoder. To be more specific, the image $\bm{I}$ is mapped to an embedding $\bm{E}(\bm{I})$ by the encoder. This embedding is then quantized based on the distance to the prototype vectors of the codebook,  $\bm{B} = \{\bm{b}_k\}$, $k \in \{1,...,K\}$. Then each embedding in $\bm{E}(\bm{I})$ is replace by the nearest prototype embedding in the codebook $\bm{B}$. We call the quantized embedding ``visual features" $\bm{Q}^n$ where $n$ is the index of the nearest prototype vector, 
$\bm{Q}^n = \bm{b}_k $, where,
\begin{equation}
n = \arg \min _{j}\left\|\bm{E}(\bm{I})-\bm{b}_{j}\right\|
\end{equation}
The set of index $n$ is called as estimated index vector, $\bm{N}$. The decoder reconstructs the image ${\bm{I}}'$ from the corresponding vectors in the codebook with another non-linear function. To learn these mappings, the gradient of the reconstruction error is then back-propagated through the decoder and the encoder using a straight-through gradient estimator \cite{bengio2013estimating}. The training objective is: 
\begin{equation}
\mathbb{L}_{E} =  \frac{\left\|\bm{I} - \bm{I}'\right\|_{2}^{2}  }{{\delta_{\bm{I}}}} + \left\|\operatorname{sg}\left[\bm{E}(\bm{I})\right]-\bm{b}_k\right\|_{2}^{2} +\beta\left\|\bm{E}(\bm{I})-\operatorname{sg}[\bm{b}_k]\right\|_{2}^{2}
\end{equation}
where $\beta$ is the loss weight and $\delta_{I}$ is the diagonal co-variance of training images and $\operatorname{sg}$ represents the stop-gradient operator that is defined as an identity at the forward computation time and has zero partial derivatives. 

The imagined visual features of $\bm{A}$ are calculated by the acoustic spectrum transformation network, which is a non-linear mapping from $\bm{A}$ to an estimated index vector  $n = \bm{G}(\bm{A})$ for $\bm{B}$. Therefore, we define the quantized vector as ``imagined visual features" $\bm{Q}^{\bm{G}(\bm{A})}$. Since the index vectors have far fewer elements than the quantized vectors, the acoustic spectrum transformation network is easy to generate, which is also verified by our experiments. To help guide the transformation network to produce imagined visual features (IVF) from audio signals that closely match visual features produced from real images, we further add a discriminator network to distinguish between the real features $\bm{Q}^{\bm{G}(\bm{A})}$  and $\bm{Q}^n$, where $n$ is calculated from $\bm{E}(\bm{I})$ via Eq. (2). The training objective is designed based on WGAN-GP \cite{gulrajani2017improved}:
\begin{equation}
\begin{aligned}
    \mathbb{L}_{D} &=\mathbb{E}[\bm{D}(\bm{Q}^n)-\bm{D}(\bm{Q}^{\bm{G}(\bm{A})})]\\ &-\lambda \mathbb{E}\left[\left(\left\|\nabla \bm{D}(\bm{C})\right\|_2 -1\right)^{2}\right] 
\end{aligned}
\end{equation}
where 
\begin{equation}
\bm{C} = w\bm{Q}^{\bm{G}(\bm{A})}+(1-w)\bm{Q}^n
\end{equation}
$\lambda$ is weight of discriminator loss, $\bm{D}$ is the discriminator with cross-entropy loss, $\nabla$ is the gradient operator and $w$ is random value from 0 to 1. $\bm{C}$ represents the penalty sampled uniformly along straight lines between pairs of points sampled from the real feature distribution and the fake feature distribution. Our full objective of transformation network is: $\mathbb{L} =  \mathbb{L}_{E} + \mathbb{L}_{D}$.

\subsection{Classifier network }
The classifier is trained on audio recordings ${\bm{A}}$ of the target dataset, such as DCASE, where the previous transformation network $\bm{G}$ and codebook $\bm{B}$ are fixed. The goal of the classifier network is to predict the category of the audio recording in the audio dataset, such as DCASE dataset. It consists of eight convolution layers, four mean-pooling and a max-pooling layer, the output dimension is the number of the category with a cross-entropy training loss.

\subsection{Implementation detail}

To save the computation cost, all frames in the training Youtube 8M dataset are resized to $512\times512$ pixels. For the audio part of Youtube 8M, DCASE and ESC-50, the audio waveform is resampled at 16k Hz. The Spectrums are computed by taking STFT with 16 FFT frequency bands. The training and test sets of DCASE include 14.4K and 5.8K audio recordings. Our network is implemented in TensorFlow and optimized by ADAM with $\beta_{1}$ = 0.5, $\beta_{2}$ = 0.0002, and the learning rate of 0.001 and the batch size of 16 for 9000 epochs. The network was trained for 223 hours on 4 NVIDIA 2080 Ti.

\section{Experimentation Results}

\subsection{Dataset}

We train our model on a subset of Youtube 8M dataset, DCASE datasets \cite{mesaros2019acoustic} and ESC-50 dataset \cite{piczak2015esc}. The subset Youtube dataset consists of ten kinds of scenes: Airport, Bus,  Street pedestrian, Park, Metro, Street traffic, Shopping mall, Public Square, Metro Station and Tram, which matches to DCASE dataset. The image size is $512* 512$. For the audio data, the length of recording is $10$ seconds for Youtube 8M and DCASE and $5$ seconds for ESC-50. The audio waveform is re-sampled at 16 kHz. Training, validation and testing data include 14K videos, 2k videos and 1k videos. The total length is 472 hours. Our network is implemented in Keras and TensorFlow and optimized by ADAM with $\beta_{1}$ = 0.5, $\beta_{2}= 0.0002$, and the learning rate of $0.0002$ and the batch size of $640$ for $5000$ epochs. The network was trained for $436$ hours on a workstation hosting $4$ NVIDIA 2080-Ti.

\subsection{Evaluate imagined visual features}

\setlength{\tabcolsep}{4pt}
\begin{table}[tbp]
\begin{center}
\caption{Acoustic Scene Classification Accuracy using different features on DCASE.}
\label{tab:DCASE2}
\begin{tabular}{llll}
\hline\noalign{\smallskip}
No. & Method  &   Seen & Unseen\\
\noalign{\smallskip}
\hline
\noalign{\smallskip}
1& CQT \cite{schorkhuber2010constant}& 72.3 & 64.6\\
2& Gammatone \cite{ellis2009gammatone} & 77.5 & 68.1\\
3& MFCC \cite{xu2004hmm} & 64.4 & 59.6 \\
4& log-mel & 73.4 & 62.5 \\
5& log-mel + delta & 76.5 & 65.1 \\
6& log-mel+deltas+delta-deltas &77.2 & 66.5 \\
7& IVF (without VQ) & 78.3 & 73.2\\
8& IVF (Q) & 81.5& 79.3\\
9& IVF & \textbf{83.7} & \textbf{82.5}\\
\hline
\end{tabular}
\end{center}
\end{table}
\setlength{\tabcolsep}{1.4pt}

To understand the influence of the acoustic spectrum transformation network of the propose classification algorithm, we compare the the visual feature with other baseline features, which is widely used in DCASE challenge. Apart from that, to show the influence of the VQ layer of the transformation model,  we modify the transformation model and evaluate the performance of the ablated model in Table \ref{tab:DCASE2}. In the table, model (1-3) shows the performance of CQT (Constant-Q transform) \cite{schorkhuber2010constant}, Gammatone \cite{ellis2009gammatone} and MFCC \cite{xu2004hmm}. Models (4-6) show the performance of the Mel family Spectrum, which gives higher accuracy than models (1-3). Model (4) uses the log-mel Spectrum, while Model (5) uses the log-mel + delta. In Model (6), the log-mel and its deltas and delta-deltas \cite{weiss2017sequence} without padding are considered. For mel Spectrum family, ``log-mel+deltas+delta-deltas'' \cite{weiss2017sequence} provides the highest accuracy. Model (7) with log-mel shows the performance of the model without the VQ layer, which decreases performance. The reason maybe that the VQ layer and generating the index can decrease the complexity of the imagined visual features. Model (8) shows the performance of the model where the output of the transformation network is the quantized embedding $\bm{Q}^n$. Model (9) shows the performance of the full proposed model where the output of the transformation network is the estimated index of codebook $\bm{B}$.
Our proposed method achieves the high accuracy on the seen and unseen environment.

\subsection{Comparison with the state-of-the-art methods}  

In this section, we compare our proposed classification method with other the state-of-the-art methods on the DCASE and ESC-50 dataset. Tables \ref{tab:DCASE} and \ref{tab:esc} show the accuracy of different methods tested on DCASE and ESC-50 dataset respectively. As the length of audio recording is different in DCASE and ESC-50, their classifiers are separately trained. 

%The input of the transformation network is ten seconds for DCASE while it is the five seconds for ESC-50. 

On both benchmarks, we convincingly beat the previous state-of-the-art methods. For DCASE, our method improves the accuracy by 7 $\%$ for the seen cities and 10 $\%$ for unseen cities, since our proposed is trained on YouTube videos recorded in different cities. The results of our proposed model and Suh et al. \cite{suh2020designing} both indicate that the GAN networks can improve the accuracy of the acoustic scene classification by capturing visual information from video datasets. For evaluating the effect of the Youtube 8M dataset, the accuracy of Suh et al. \cite{suh2020designing} trained on a labelled version Youtube 8M and DCASE (Suh et al. \cite{suh2020designing} + 8M) is 82.4 for seen cities and 76.2 for unseen cities, which is lower than that of our method. Youtube 8M dataset has some wrong labels data, which may affect the performance of other state of the art methods. It is important to note that our transformation network is trained without label information. The accuracy is further improved 8\% by fine-tuning with a target dataset.

For ESC-50, we have 2$\%$ higher accuracy than human performance and have a marginally better performance to the transfer learning method \cite{kumar2018knowledge}. These results show that a large number of video data can improve the performance of classification. Note that since the size of the Youtube 8M dataset is huge, we only train our network with a subset for efficiency, so it is possible that further gains can be achieved by using all the available training data.

\setlength{\tabcolsep}{4pt}
\begin{table}[tbp]
\begin{center}
\caption{Acoustic Scene Classification Accuracy at the seen cities and unseen cities on DCASE.}
\label{tab:DCASE}
\begin{tabular}{lll}
\hline\noalign{\smallskip}
Method  &   Seen Cities & Unseen Cities\\
\noalign{\smallskip}
\hline
\noalign{\smallskip}
SoundNet \cite{aytar2016soundnet}  & 75.8 &  67.5 \\
Gao et al. \cite{gaoacoustic} &	77.0 & 73.1 \\
Hu et al. \cite{hu2020device} & 77.5 & 74.7 \\
Suh et al. \cite{suh2020designing} & 78.1 & 74.6 \\ 
Suh et al.+ 8M & 82.4 & 76.2 \\ 
Ours &  \textbf{83.7} & \textbf{82.5} \\
\hline
\end{tabular}
\end{center}

\end{table}
\setlength{\tabcolsep}{1.4pt}

\setlength{\tabcolsep}{4pt}
\begin{table}[tbp]
\begin{center}
\caption{Acoustic Scene Classification Accuracy on ESC-50.}
\label{tab:esc}
\begin{tabular}{lll}
\hline\noalign{\smallskip}
Method & Accuracy\\
\noalign{\smallskip}
\hline
\noalign{\smallskip}

SVM-MFCC \cite{piczak2015esc} & 39.6\\
Autoencoder \cite{aytar2016soundnet} & 39.9\\
Random Forest  \cite{piczak2015esc} & 44.3\\
Piczak ConvNet  \cite{piczak2015environmental} & 64.5\\
SoundNet \cite{aytar2016soundnet} & 74.2\\
AV learning \cite{arandjelovic2017look}  & 79.3\\
Human \cite{piczak2015esc} & 81.3 \\
Transfer learning \cite{kumar2018knowledge} & 83.5\\
Ours & \textbf{83.7} \\

\hline
\end{tabular}
\end{center}
\end{table}
\setlength{\tabcolsep}{1.4pt}

\section{Conclusion}

We have proposed a spectrum transformation network in order to address the unseen cities problem in the scene classification task. The proposed network is able to learn the correlation between audio and visual features and transform a log-mel signal to imagined visual features. The imagined visual features are used to classify audio recordings as other traditional spectra. The transformation network can be trained on unlabelled datasets while the classifier only needs a limited number of target data. The experiments on DCASE dataset and ESC-50 dataset show we can achieve higher accuracy than the state-of-the-art methods, especially for unseen cities. Our future research will focus on reconstructing scenes with detailed geometric features from audio recordings.

\bibliographystyle{IEEEbib}
\bibliography{strings,refs}

\end{document}